\documentclass[journal,article,submit,pdftex,moreauthors]{Definitions/mdpi}
\usepackage[linesnumbered,ruled]{algorithm2e}
\firstpage{1} 
\pubvolume{1}
\issuenum{1}
\articlenumber{0}
\pubyear{2022}
\copyrightyear{2022}
\datereceived{} 
\dateaccepted{} 
\datepublished{} 
\usepackage{changes}
\definechangesauthor[name=Bartosz, color=red]{BD}

\hreflink{https://doi.org/} 
\pdfoutput=1
\Title{A method to explore flavor symmetries of the 3HDM and their implications on lepton masses and mixing}

\TitleCitation{A method to explore flavor symmetries of the 3HDM and their implications on lepton masses and mixing}

\Author{Bartosz Dziewit$^1$*, Marek Zra\l ek$^2$, Joris Vergeest $^1$*  and Piotr Chaber$^1$}

\AuthorCitation{Dziewit, B.; Zra\l ek, M.; Vergeest, J.; Chaber, P.}

\address{%
$^{1}$ \quad Institute of Physics, University of Silesia, 75 Pu\l{}ku Piechoty 1, 41-500 Chorz\'{o}w, Poland\\
$^{2}$ \quad Humanitas University in Sosnowiec, ul. Kilińskiego 43, 41-200~ Sosnowiec, Poland}

\corres{Correspondence: BD: bartosz.dziewit@us.edu.pl;
JV: jorisvergeest@hotmail.com}

\abstract{We present a method to identify symmetry groups of the Yukawa sector of the three-Higgs-doublet model and to determine the implication that the symmetry has on the lepton masses and mixing. The method can accommodate different hypotheses about the group representation assignments, and thus support the exploration of candidate symmetry groups. For one particular representation selection scheme we apply the computer-implemented method to scan all discrete groups of order less than 1035. It can be proven that none of these groups defines a flavor symmetry that implies masses and neutrino mixing angles consistent with the experimental lepton data, although several cases are found that are partially or approximately consistent.}

\begin{document}
\nolinenumbers
\section{Introduction}

Many important questions still remain unanswered in the Standard Model (SM), which on the other hand agrees with experiment with great accuracy \cite{Harari:1977kv,IBE2016365}. In this situation, various extensions of the SM have been proposed to address some unresolved questions, for example, the origin of the three generations of fermions, with such different mass hierarchies and radically different mixing patterns of leptons compared to quarks. In the lepton sector, which is the focus of this work, various attempts have been made to devise a theory that could predict the lepton masses and their mixing (e.g. see \cite{King:2014nza,Rong:2019ubh,Joshipura:2016quv, Qian:2015waa,Frampton:1994rk}). The experimental data on these quantities have gained better precision in recent years (e.g. see \cite{Aker:2022ijt, T2K:2019bcf, RENO:2018dro}), and it is getting increasingly hard to explain them satisfactorily. An often encountered approach to this problem is to look for some flavor symmetry of the lepton interaction Lagrangian that can, in a way consistent with the experimental data, predict the values of the lepton masses and their mixing angles. It is well known that a continuous nontrivial flavor symmetry in the lepton sector does not exist, which follows from the fact that the lepton masses are distinct \cite{Feruglio:2019ybq}. Discrete symmetries remain possible in the Yukawa sector. However, due to Schur’s first lemma, the acting of two inequivalent representations in flavor space (one on the lepton doublets, one on the lepton singlets) implies that any mass matrix is either proportional to the unit matrix or vanishes. One strategy to avoid this problem is to break the flavor symmetry explicitly in such a way that the charged lepton mass matrix and the neutrino mass matrix are separately invariant under two different subgroups of a larger discrete symmetry group $G$ \cite{deAdelhartToorop:2011re}. The subgroups are usually kept small, although in \cite{Holthausen:2012wt}, groups $G$ of order up to 1000 have been investigated. It was shown that the possible lepton mixing patterns then depend on how the two subgroups are embedded within $G$. Another approach is based on a Lagrangian including mass terms constructed with lepton, Higgs doublet and scalar flavon fields. Group-invariance of the terms is looked for by systematic probing all plausible representation assignments \cite{tribimaximalsmall}, where dynamic parameters (VEVs) affect the predicted mixing angles \cite{Lam:2011ag} \cite{lam:2012ga}.

The aforementioned explicit flavor symmetry breakdown in the Yukawa sector can be avoided in the presence of Higgs fields (causing spontaneous symmetry breaking) that transform under $G$. Two-Higgs doublet models (2HDM) \cite{BRANCO20121} are obvious candidates to investigate that principle. In \cite{Chaber:2018cbi} and \cite{sym12010156} non-Abelian groups that have two- and three-dimensional irreducible representations have been investigated. The lepton masses and the angles of the mixing matrix resulting from the assumed discrete symmetry were found, but the agreement with the experimental data was not satisfactory. Although the implied lepton masses of the model were nondegenerate, the obtained mixing matrices appeared to be always monomial. However, the results are different when one more Higgs doublet is added, as for example in the three-Higgs-doublet model (3HDM) \cite{Grossman:1994jb, Keus:2013hya}. This model contains neither flavons nor other additional fields and it allows the treatment of neutrinos as Dirac particles or as Majorana particles. We will assume that the Higgs doublets form a flavor vector transforming under a group $G$, as do the charged-lepton and neutrino flavor vectors. In this model (\cite{Chaber:2018cbi}, \cite{Ivanov_2013}), the mass-squared matrices are not affected by Schur’s first lemma: the masses can be nondegenerate and the neutrino mixing nontrivial, even when the assigned 3D irreps are inequivalent and $G$ is non-Abelian. 

The 3HDM symmetry depends on how group representations are assigned to the flavor vectors. In order to either prove the existence of a proper symmetry or to exclude any symmetry, different hypotheses about the representation assignments must be tested. For this purpose we have built a computer tool that analyses discrete groups of arbitrary order. The theoretical foundation, the design of the tool and the initial outcomes are presented in this article.

In the following section we define the extension of the SM to 3HDM and describe the construction of the Yukawa Lagrangian. In Chapter 3 we define one particular scheme for the representation assignments and summarize the scan results for groups with order of less than 1035. Chapter 4 describes the method for finding those transformations (defined by finite subgroups of $U(3)$) that impose non-trivial relationships between lepton masses and mixing angles. In Chapter 5 we list alternative assignments of the representations, and directions for improvement of the exploration method.

\section{Flavor symmetry of the 3HDM}
The SM comprises one electroweak $SU(2)$ Higgs doublet, that generates the masses of the gauge bosons and the fermions.
The fermions that we study in this work are the three charged leptons electron, muon and tau, denoted $l_e$, $l_\mu$ and $l_\tau$, and three neutrinos, that we name $\nu_e$, $\nu_\mu$ and $\nu_\tau$. The subscripts $e$, $\mu$ and $\tau$ are called lepton flavors.

Suppose for a moment that the three charged leptons have equal mass and also the three neutrinos have equal mass. Then the charged leptons would be indistinguishable and the same goes for the neutrinos. Any Lagrangian constructed from the lepton fields would be invariant under interchanging of charged fermions or interchanging neutrinos or, more generally, under any unitary transformation of the triplets (or flavor vectors) $(l_e, l_\mu,l_\tau)^T$ and $(\nu_e$, $\nu_\mu, \nu_\tau)^T$. We would have a trivial continuous $U(3)$ flavor symmetry. If any two charged leptons (or any two neutrinos) have different masses, the Lagrangian will in general lose its $U(3)$ flavor symmetry. We will see that for the Lagrangian of the SM we can apply the reverse argument: if the Yukawa term is invariant under certain continuous $U(3)$ transformations of the flavor vectors then the masses of the charged leptons, and of the neutrinos, are degenerate. The key point here is that unlike the SM the 3HDM does allow nondegenerate lepton masses under specific discrete symmetries and, in addition, predict relations among mass and mixing quantities that can be compared against the experimental data.

In the following we consider two scenarios. In the first we treat all six fermions as Dirac particles. In the second we treat the three neutrinos as Majorana particles instead. A difference between the Dirac neutrinos and Majorana neutrinos is the way they acquire mass through the Higgs mechanism. As an extension of the SM the 3HDM has three electroweak SU(2) Higgs doublets (instead of only one), denoted $\Phi_i$, $i$=1,2,3. Doublet $i$ contains an electrically neutral Higgs field $\phi_i^0$ and a negative Higgs field $\phi^-_i$. The left-handed components of the leptons form electroweak SU(2) doublets $L_\alpha = (\nu_{\alpha L}, l_{\alpha L})^T$ for each of three flavors $\alpha$. Lepton masses arise from Yukawa couplings between the Higgs fields and the (left- and right-handed components of the) lepton fields.

To define flavor symmetry we first need to know the Lagrangian of the system responsible for the lepton masses and mixing. In this work we account for the Yukawa sector of the Lagrangian only, not for the vacuum alignments imposed by symmetry of the potential energy term of the Higgs bosons.
The Yukawa Lagrangian $\mathcal{L}^l$ of the charged leptons has 27 terms (plus their Hermitian conjugates), for the occurring combinations $(i,\alpha,\beta)$, compactly denoted 
\begin{eqnarray} \label{LAG1}
\mathcal{L}^l = - (h^l_i)_{\alpha \beta} \overline{L}_{\alpha L} \tilde{\Phi}_i l_{\beta R} + \text{H.c.}\label{Yukawa1}.
\end{eqnarray}
Here summation over the lepton flavor indices $\alpha, \beta = e, \mu, \tau$ and over $i=1,2,3$ is understood. The Higgs doublet $\tilde{\Phi_i}= i\sigma_2\Phi_i^\ast$ is the complex conjugate representation of $\Phi_i$, where $\sigma_2$ is the second Pauli matrix.
$h_i$ is the $3\times3$ Yukawa matrix defining the couplings due to Higgs doublet $\Phi_i$.
$\overline{L}_{\alpha L} =
(\overline{\nu}_{\alpha L},\overline{l}_{\alpha L})$.
If we treat neutrinos as Dirac particles, i.e. there exist right-handed neutrino fields, then the Yukawa Lagrangian of the neutrinos is (similar to $\mathcal{L}^l$):
\begin{eqnarray}
\mathcal{L}^{\nu} = - (h^\nu_i)_{\alpha \beta} \overline{L}_{\alpha L} {\Phi}_i \nu_{\beta R}+ \text{H.c.}
\end{eqnarray}
If we assume that neutrinos are Majorana particles instead then separate right-handed neutrino fields are not needed. For each flavor $\alpha$ the charge-conjugate lepton doublet $L^c_{\alpha L} =
C\overline{L}^T_{\alpha L}$ is a right-handed doublet. Then an effective dimension-five operator can be constructed to define the Yukawa Lagrangian of Majorana neutrinos as \cite{Weinberg:1979sa}
\begin{eqnarray}
\mathcal{L}^M =-\frac{g}{M} (h^M_{i j})_{\alpha \beta}(\overline{L}_{\alpha L} \Phi_i)(\Phi^T_j L^c_{\beta L})+ \text{H.c.,}
\end{eqnarray}
where $g$ and $M$ are constants and $h^M_{ij}$ is a $3\times3$ matrix of coupling constants. The number of terms of this Lagrangian is 81 (not counting the H.c. terms), which is the number of combinations $(i,j,\alpha,\beta)$.

After spontaneous electroweak symmetry breaking, the vacuum expectation values $v_i$ for the $\Phi_i$ generate the mass matrices $M^l$, $M^{\nu}$ and $M^M$ as
\begin{eqnarray}\label{massmatrix}
M^l  =-\frac{1}{\sqrt2} v^*_ih^l_i  \hspace{0.5cm}
M^{\nu}  =\frac{1}{\sqrt2}v_ih^{\nu}_i \hspace{0.5cm}
M^M  =\frac{g}{M} v_iv_jh^M_{ij}.
\end{eqnarray}
The mass-Lagrangian terms for the charged leptons, and for the Dirac neutrinos and the Majorana neutrinos will read
\begin{eqnarray} \label{LAG2}
\mathcal{L}^l_{mass} =  - \overline{l}_LM^ll_R  + \text{H.c.}\hspace{0.5cm}
\mathcal{L}^\nu_{mass} =  - \overline{\nu_L}M^{\nu}\nu_R  + \text{H.c.}\hspace{0.5cm}
\mathcal{L}^M_{mass} =  - \frac{1}{2}\overline{\nu_L}M^M\nu_L^c + \text{H.c.},
\end{eqnarray}
where $l_L$, $l_R$, $\nu_L$, $\nu_R$ are flavor vectors for the left/right-handed charged leptons and neutrinos, respectively. For example $l_L = (l_{eL},l_{\mu L},l_{\tau L})^T$. We also define $v=(v_1,v_2,v_3)^T$ as the flavor vector of the Higgs doublets.

We are now ready to explore symmetries of the three Lagrangian terms; for that purpose we can leave out the Hermitian conjugate terms, so it is sufficient to consider the quantities
\begin{eqnarray} \label{quant}
\overline{l}_L \, (v_i^* h^l_i) \, l_R, \, \,\hspace{0.5cm}
\overline{\nu_L}\, ( v_i h^\nu_i) \, \nu_R \, \,\hspace{0.5cm} 
\text{and} \, \,\hspace{0.5cm}
\overline{\nu_L} \, (v_i v_j h^M_{ij}) \, \nu_L^c.
\end{eqnarray}
$l_L$, $l_R$, $\nu_L$, $\nu_R$ and $v$ will each be assigned a three-dimensional representation of some finite group $G$. In the present study we restrict to irreducible representations; other choices of representations will be discussed later. All representation matrices must be unitary in order to conserve the total lepton number and to ensure that $\sqrt{\Sigma |v_i|^2}=(\sqrt{2}G_F)^{-1/2}$=246 GeV, where $G_F$ is the Fermi coupling constant. We should remember that $l_l$ and $\nu_l$ reside in the same SU(2) doublet and hence should be assigned the same representation.

As said in the introduction, we want to find such a discrete symmetry of the full Lagrangian of leptons that will give mass matrices of charged leptons and neutrinos, where several free parameters can be selected so, that after diagonalization, we obtain lepton masses, mixing angles and the CP symmetry  breaking phases consistent with experimental observations. For this purpose, we should study the symmetry of two terms, 1) the Yukawa Lagrangian which, after spontaneous symmetry breaking, produces lepton mass matrices, and 2) the Higgs potential. In the present study, we will only deal with the symmetry of a well-defined Yukawa Lagrangian, leaving for further analysis the selection of a Higgs potential the form of which is more freely defined.

The first step is to identify distinct groups $G$, isomorphic to a $U(3)$ subgroup, that have one or more unitary three-dimensional irreducible representations, and assign these representations to the flavor vectors so that some or all of the three transformed terms of Eq. \eqref{quant}
\begin{eqnarray} \label{terms1}
&& \overline{l}_L A_L^\dagger(g)\, (A_\Phi^*(g) v^*)_i h^l_i\, A_{l R}(g) \, l_R \\
&& \overline{\nu}_L A_L^\dagger(g)\, (A_\Phi(g) v)_i h^\nu_i\, A_{\nu R}(g) \, \nu_R \label{terms2}\\
&& \overline{\nu}_L A_L^\dagger(g)\,(A_\Phi(g) v)_i (A_\Phi(g) v)_j) h^M_{ij} \, A^*_L(g) \, \nu^c_L \label{terms3}
\end{eqnarray}
remain invariant under simultaneous operation of matrices of the irreducible representations $A_L(g), A_{lR}(g), A_{\nu R}(g)$ and $A_\Phi(g)$ for all $g$ in $G$ and any vector $l_L$, $l_R$ and $v$. Note that different flavor vectors can be assigned different representations of $G$.

As an example let us consider the Yukawa Lagrangian for the charged leptons \eqref{terms1}. It is $G$-invariant if
\begin{eqnarray}
\overline{l}_L \, (v_i^* h^l_i) \, l_R = \overline{l}_L A_L^\dagger(g)\, (A_\Phi^*(g) v^*)_i h^l_i\, A_{l R}(g) \, l_R, \hspace{0.5cm} \text{all} \, g\in G,
\, \text{all} \,\, l_L, l_R, v. \label{operatoreq1}
\end{eqnarray}
 Since all matrix operators are linear, we are dealing with a set of $|G|$ equations with 27 unknown complex numbers (the entries of the three Yukawa matrices $h^l_i$). $|G|$ is the order of $G$. So if Eq. \eqref{operatoreq1} can be solved, it defines the $h^l_i$. Here we realize that if Eq. \eqref{operatoreq1} is solved for a particular group element $g$ then the system of equations obtained after transformation of all flavor vectors with the operators $A_L(g), A_{lR}(g), A_{\nu R}(g)$ and $A_\Phi(g)$, will also be solvable for $g$, giving exactly the same solution(s) $h^l_i$. So if Eq. \eqref{operatoreq1} is solved for $g$, it will be solved for $g^n$ for all $n$. In that case we can remove the mutually equivalent equations. Further, if Eq. \eqref{operatoreq1} is solved for group elements $g$ and $h$ it will be solvable for $g^n \, h^m$ too. More generally, we need only to take into account the generators of $G$, rather than all its elements.
 In a similar way we can define the set of equations to solve for $h_i^\nu$ and $h^M_{ij}$:
 \begin{eqnarray}
\overline{\nu}_L \, (v_i h^\nu_i) \, \nu_R =&&\overline{\nu}_L A_L^\dagger(g)\, (A_\Phi^*(g) v)_i h^\nu_i\, A_{\nu R}(g) \, \nu_R, \label{operatoreq2} \\
\overline{\nu}_L (v_i v_j)h^M_{ij} \nu_L^c =&&\overline{\nu}_L A_L^\dagger(g)\,(A_\Phi(g) v)_i (A_\Phi(g) v)_j) h^M_{ij} \,
A^*_L(g) \, \nu^c_L \label{operatoreq3},
\end{eqnarray} 
where for Eq. \eqref{operatoreq3} the number of unknowns is 81 (the entries of nine $3\times3$ matrices).
A discrete symmetry of lepton interactions given by some group G requires that Eqs. \eqref{operatoreq1} \textit{and} \eqref{operatoreq2} are satisfied for Dirac neutrinos and Eqs. \eqref{operatoreq1} \textit{and} \eqref{operatoreq3} for Majorana neutrinos for all generators of G.
 
\section{\label{sec:3}Solving the invariance equations}
We take Eq. \eqref{operatoreq1} as an example again. Working out the matrix multiplications it can be expressed as
\begin{equation}\label{kron1}
    ((A_\Phi(g))^\dagger\,\otimes\,(A_L(g))^\dagger\,\otimes\,
    (A_{lR}(g))^T)\,h^l\,=\,h^l,
\end{equation}
The Kronecker product gives a 27$\times$27 matrix, and $h^l$ is the 27-dimensional vector built from the Yukawa matrices $h^l_1, h^l_2$ and $h^l_3$, in that order, row-wise. If $h^l$ is an invariant eigenvector satisfying Eq.\eqref{kron1} for all generators of $G$ then $\mathcal{L}^l$ is $G$-invariant. The invariance equations for the terms $\mathcal{L}^\nu$ and $\mathcal{L}^M$ are
\begin{eqnarray}
   ((A_\Phi)^T\otimes(A_L)^\dagger\otimes
    (A_{\nu R})^T) h^\nu && =  h^\nu \label{kron2}\\
    ((A_\Phi)^T\otimes(A_\Phi)^T\otimes(A_L)^\dagger
    \otimes(A_L)^\dagger) h^M && =  h^M \label{kron3},
\end{eqnarray}
dropping the $(g)$-argument for clearness. $h^\nu$ is a 27-dimensional invariant eigenvector built from the entries of the three $h^\nu_i$ matrices, row-wise. The 81-dimensional vector $h^M$ contains the entries of $h^M_{11}$, row-wise, followed by the entries of $h^M_{12}$, $h^M_{13}$, $h^M_{21}$ etc.

For completeness we note that
Eqs. \eqref{kron1}, \eqref{kron2} and \eqref{kron3} are equivalent to the (Clebsch-Gordan) tensor product decomposition equations \cite{ludl}
\begin{eqnarray}
    A_L \otimes A_{lR}^* \, && =  A_\Phi^* \oplus ...\label{decomp1} \\
    A_L \otimes A_{\nu R}^* \, && =  A_\Phi \oplus ... \label{decomp2} \\
    A_\Phi^* \otimes A_L \otimes A_L \, && =  A_\Phi \oplus ... \label{decomp3},
\end{eqnarray}
still assuming that all matrix operators are unitary.
Eqs. \eqref{decomp1} (or \eqref{decomp2}) are fulfilled if and only if the 9-dimensional tensor product operator allows a decomposition containing at least one three-dimensional matrix operator. Eq. \eqref{decomp3} requires that the 27-dimensional tensor product operator contains at least one three-dimensional matrix operator.

To investigate which groups and which representation assignments can accomplish such invariance one could expect that it is necessary to explicitly solve the invariance equations for each instance. However, in the next section it will be shown that a significant part of those calculations can be skipped. The computational complexity of the calculations needs to be considered further because
groups with $|G|> \sim 1000$ can have 8 or more generators and hundreds of inequivalent three-dimensional representations, leading to up to 10 million equation sets Eqs. \eqref{kron1}-\eqref{kron3} for a single group. We therefore put some restrictions to the analysis. Only irreducible representations are considered and a candidate group must have at least one faithful three-dimensional representation. Our precise selection criterion is that at least one of the irreducible representations assigned to a mass term is faithful. As mentioned, different selection criteria can be tested too.
 
 From a scan of groups with $|G|$<1035, while applying the selection criteria, we find 749 groups that provide one or more solutions $h^l$ to Eq. \eqref{kron1}, and in each of these cases also a solution $h^\nu$ to Eq. \eqref{kron2}. In the entire set of (over 6 million) solutions $h^l$ (or $h^\nu$), 2130 are linearly independent. 216 groups provide one or more solutions to Eq. \eqref{kron3}, resulting in 70 linearly independent vectors $h^M$.

It follows from Eq. \eqref{massmatrix} that for given solutions $h^l$, $h^\nu$ and $h^M$ the mass matrices (and hence the mixing matrices) are functions of the VEVs. Since we do not know the absolute scale of the Higgs couplings nor the flavor assignments we have $v2/v1$ and $v3/v1$ as 4 real parameters, and we can at most determine mass ratios $m_i/m_j$ of the leptons. Also any PMNS matrix can be determined up to permutation of its rows and columns and (for the Majorana case) up to a phase for two rows.

The low number of solutions that, despite the 4 parameters, imply nontrivial neutrino mixing is somewhat surprising. The majority of the obtained mixing matrices is monomial and exclude mixing. For none of the tested groups we find a symmetry that is entirely compatible simultaneously with the experimental masses and mixing data, neither when the neutrinos are assumed Dirac or Majorana particles. However, there are symmetries (e.g. for $\Delta(96)$ and $S_4$) that imply partially or approximately correct quantities.
 
\section{Method to find relations among the lepton mass and mixing parameters}
The selection and processing of groups is fully automated using the computer-algebra system GAP \cite{GAP4}. To determine which representation assignments would solve a particular invariance equation it is sufficient to observe the group's character table, which is readily provided by GAP. By doing so a tremendous amount of redundant computation is avoided; we therefore go into some detail of this point.
The character of $g\in G$ in representation $A$, denoted $\chi^A (g)$, is defined as the trace of matrix $A(g)$. The mapping $\chi^A$ is called the character of $A$. Let $A$ and $B$ be representations of $G$, then
\begin{equation}
\label{innerchar}
\langle \chi^A,\chi^B \rangle := \frac{1}{|G|} \sum_{g \in G}\chi^A (g)^\star \chi^B (g)
\end{equation}
defines the inner product of characters $\chi^A$ and $\chi^B$. It can be proven that $\chi ^{A \otimes B}(g) = \chi^A(g) \chi^B(g)$ for all $g \in G$. Let also $H$ be an irreducible representation of $G$. Then $\langle \chi^{A \otimes B},\chi^H \rangle$  is the multiplicity of $H$ occurring in the tensor product decomposition of $A \otimes B$. In Eq. \eqref{decomp1},
$\langle \chi^{A_L \otimes A_{lR}^*},\chi^{A_\Phi^*} \rangle$
can take a value from 0 to 3. This is the number of linearly independent solutions $h^l$ to Eq. \eqref{kron1}. The inner product can be directly deduced from the character table of $G$, and thus prior to the actual generation of the representation matrices themselves and without explicitly solving Eq. \eqref{kron1}. For brevity let us denote the representations appearing in Eq. \eqref{decomp1} as $A$, $B$ and $H$, respectively. Then, if $A$, $B$ and $H$ are irreducible, Eq. \eqref{kron1} has a nontrivial solution if and only if
$n :=\langle\chi^{A \otimes B^\star},\chi^{H^\star} \rangle > 0$. In the present selection procedure the representation triplet $(A,B,H)$ is accepted only if $n=1$; as a trade-off regarding computational load, we disregard multidimensional solutions ($n>1$).
\begin{algorithm}
\caption{Obtain Dirac mass matrices from group symmetry of the 3HDM}
\label{alg:alg1}
 Let $S$ be a set of finite groups of interest \;
 \For{every $G$ in $S$ }{\If{3 divides the order of $G$}{
 obtain character table of $G$ \;
 obtain $E$, the set of characters for three-dimensional irreducible representations of $G$ \;
 \For{every 4-tuple of characters $(A,B,C,H)$ in $E^4$}{
  \If{one of ${A,B,H}$ is faithful
  and $n_1$:=innerProduct($A\otimes B^\ast,H^\ast$)=1}{
  Obtain unitary representation matrices for the 3D irreps of $A, B, H$ \;
 \For{every generator of $G$}{setup Kronecker product equation;}
 Solve set of Kronecker product equations for $h^l$-matrices\;
 Calculate charged lepton mass matrix $M^l$\;}
  \If{one of ${A,C,H}$ is faithful
  and $n_2$:=innerProduct($A\otimes C^\ast ,H$)=1}{
   Obtain unitary representation matrices for the 3D irreps of $A, C, H$ \;
 \For{every generator of $G$}{setup Kronecker product equation;}
 Solve set of Kronecker product equations for $h^\nu$-matrices\;
 Calculate Dirac neutrino mass matrix $M^\nu$\;}
  \If{$n_1$=1 and $n_2=1$ and one of ${A,B,H}$ is faithful\\ and one of ${A,C,H}$ is faithful\\}{
  Calculate $U_{PMNS}$ from $M^l$ and $M^\nu$;}
 }
 }
 } 
\end{algorithm}

The process of identifying 3HDM symmetry groups is schematically given in Algorithm \ref{alg:alg1}. 
The first if-statement filters out the groups with $|G|$ not divisible by 3. It can be proven that groups with 3D irreps (and those are the ones we consider) must have order divisible by 3. In the next statement the character table of $G$ is retrieved, using GAP. Let us take group $S_4$ (the group of all permutations of 4 objects) as an example. $S_4$ has order $|G|=24$. Its character table is shown in Table \ref{tab:table1}.
The top row in the table lists how many group elements are member of a particular conjugacy class of the group. $S_4$ has 5 conjugacy classes and the list adds up to 24. The first conjugacy class has one member, which is the identity element of the group. The next 5 rows list the vectors of characters for all irreducible representations of $S_4$. Each row defines the 5 entries of a character $\chi^i$, where each entry equals the trace of a representation matrix. For example $\chi^3 = (2,0,-1,2,0)$. Its 4th component specifies that the three group elements in the 4th conjugacy class each correspond to a representation matrix with trace equal to 2. The first numerical column (belonging to the identity element) indicates the dimensions of the matrix representations, since the trace of any identity matrix equals its dimension. 

In the character table of $S_4$ the algorithm finds two three-dimensional irreducible characters namely $\chi^4$ and $\chi^5$, listed in the last two lines of \ref{tab:table1}. So the set $E$ (introduced in line 5 of the algorithm) is going to contain these two characters. The algorithm also signifies that both are faithful, since all their character entries (except the first) are smaller than 3, and thus none of the corresponding representation matrices will be an identity matrix. This guarantees that any two different group elements will be represented by two different matrices. 

\begin{table*}[h!]
\caption{Irreducible characters of group $S_4$}
\label{tab:table1}
\begin{tabular}{c|ccccc|}
 & 1 & 6 & 8 & 3 & 6\\
\hline
$\chi^1$ & 1 & 1 & 1 & 1 & 1\\
$\chi^2$ & 1 & -1 & 1 & 1 & -1\\
$\chi^3$ & 2 & 0 & -1 & 2 & 0\\
$\chi^4$ & 3 & -1 & 0 & -1 & 1\\
$\chi^5$ & 3 & 1 & 0 & -1 & -1\\
\hline
\end{tabular}
\end{table*}

From step 6 of the algorithm all possible 4-tuples that can be made from the characters in $E$ are processed. What we describe from here on is for one 4-tuple of characters, named $(A,B,C,H)$. In line 7 the inner product $n_1$ determines whether decomposition Eq. \ref{decomp1} is feasible for characters $(A,B,H)$. In case $A=\chi^4$ and $B=\chi^4$ it is found that
$A\otimes B^\star =(9,-6,0,3,-6)$. For $H=\chi^4$ it follows from Eq. \eqref{innerchar} that $\langle A \otimes B^\star, H^\star \rangle = n_1 = 1$. So $(A,B,H)=(\chi^4,\chi^4,\chi^4)$ is a candidate character triplet to define a symmetry of the charged-lepton mass term. One more  triplet, $(A,B,H)=(\chi^4,\chi^4,\chi^5)$ appears viable too. As a result, only these two triplets for the charged-lepton term need further processing, in lines 8 to 13. The explicit three-dimensional matrix representations (denoted $\mathbf{3}_A,\mathbf{3}_B$ and $\mathbf{3}_H$) of $A$, $B$ and $H$ are obtained using the Repsn package of GAP \cite{Repsn}. The Kronecker product Eq. \eqref{kron1} is constructed for each generator $f_i$ of $G$, and the resulting set of equations is solved for $h^l$, using the BaseFixedSpace function of GAP. If we define the 27-dimensional matrix $K^l_i$ as
\begin{equation}\label{Kmatrix}
    K^l_i=(A_\Phi(f_i))^\dagger\,\otimes\,(A_L(f_i))^\dagger\,\otimes\,
    (A_{lR}(f_i))^T,
\end{equation}
where $i=1,\dots,n_G$ and $n_G$ is the number of generators, then BaseFixedSpace takes the matrices $K^l_i$ as input and returns their $n_1$-dimensional common eigenspace. In our application $n_1$=1, so the eigenspace defines $h^l$.

In line 13 $M^l$ is calculated as a function of $v$, using Eq. \eqref{massmatrix}. From line 15 on it is checked whether assignment $(A,C,H)$ can solve decomposition Eq. \eqref{decomp2} and if so, the Dirac neutrino mass matrix $M^\nu$ is calculated. Finally, if both $M^l$ and $M^\nu$ are available, $U_{PMNS}$ is calculated (using Mathematica \cite{Mathematica}) for the duplet $((A,B,H),(A,C,H))$ using the corresponding representations $((\mathbf{3}_A, \mathbf{3}_B, \mathbf{3}_H),(\mathbf{3}_A, \mathbf{3}_C, \mathbf{3}_H))$. For group $S_4$ it turns out that an invariant $U_{PMNS}$ exists, as will be described later.

For a given $(A,B,C,H)$, Algorithm \ref{alg:alg1} outputs either nothing or $M^l$ or $M^\nu$ or both mass matrices, and in the latter case $U_{PMNS}$ as well. From this output the invariance of the individual terms $\mathcal{L}^l$ or $\mathcal{L}^\nu$ can be studied, as well as the invariance of ($\mathcal{L}^l$+$\mathcal{L}^\nu$). Since $U_{PMNS}$ is extracted from observations of $SU(2)_L$-symmetric charged current interactions the characters $A$ and $H$ both appear twice in duplet $((A,B,H),(A,C,H))$; $l_{\alpha L}$ and $\nu_{\alpha L}$ both reside in the same $SU(2)_L$ doublet and thus $l_L$ and $\nu_L$ should transform equally. Likewise the transformations of $v$ and $v^\star$ differ by complex conjugation, since the two corresponding Higgs doublets differ by complex conjugation.

In case the neutrinos have Majorana nature, the computations are similar to Algorithm \ref{alg:alg1}. The character assignments are then of the form $((A,B,H),(A,H))$.

\begin{figure}[H]
\begin{adjustwidth}{-\extralength}{0cm}
\centering
\includegraphics[width=18 cm]{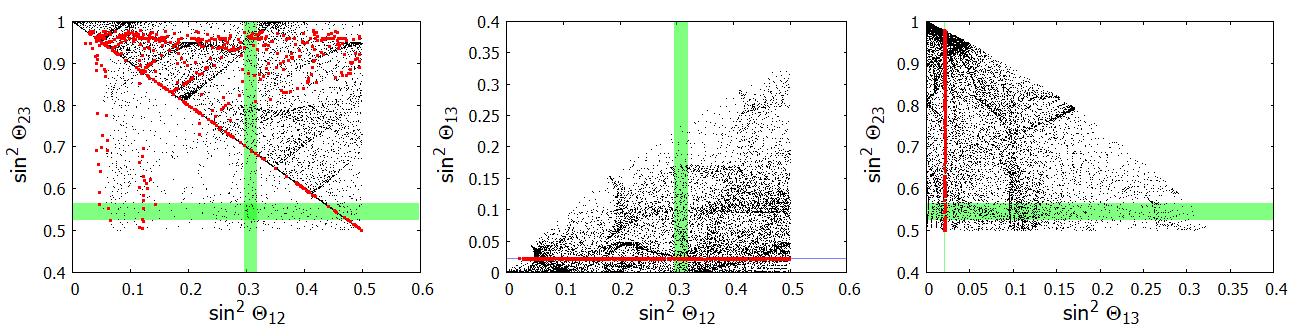}
\end{adjustwidth}
\caption{\label{dmix}$U_{PMNS}$ mixing angles for $S_4$-symmetry; $|v_i/v_1| \in [0, 1.2]$. The shaded areas indicate $1\sigma$ intervals of the quantities. The red points deviate less than $1\sigma$ from the experimental values of sin$^2 \theta_{13}$ and $\delta_{CP}$.
\label{fig:dmix}}
\end{figure}

From $M^l$, $M^\nu$ (or $M^M$) and $U_{PMNS}$ the 3HDM predicts lepton mass ratios and mixing angles as functions of $v_2/v_1$ and $v_3/v_1$. As an example we take group $S_4$ again, which happens to be the smallest group allowing nontrivial flavor mixing in case neutrinos are Dirac particles. The two representation assignments that provide such neutrino mixing symmetry are 
$((\mathbf{3}_1,\mathbf{3}_1,\mathbf{3}_1),\,(\mathbf{3}_1,\mathbf{3}_1,\mathbf{3}_1))$ and
$((\mathbf{3}_1,\mathbf{3}_1,\mathbf{3}_2),\,(\mathbf{3}_1,\mathbf{3}_1,\mathbf{3}_2))$, where $\mathbf{3}_1$ and $\mathbf{3}_2$ are the two inequivalent three-dimensional representations of $S_4$. The first representation assignment leads to anti-symmetric mass matrices and a monomial mixing matrix. The second implies the mass matrices:
\begin{equation}
\begin{array}{cc}
M^l = -\frac{c_l}{\sqrt2}
\begin{pmatrix}
0       &   v_3^\star        &      v_2^\star     \\
v_3^\star  &   0        &      v_1^\star     \\
v_2^\star       &   v_1^\star        &      0     
\end{pmatrix},
 M^{\nu} = -\frac{c_\nu}{c_l} M^{l*},
\end{array}
\end{equation}
where $c_\nu$ is an arbitrary constant. 
Based on these two mass matrices, we obtain fits to the experimental neutrino mixing data with sin$^2\theta_{12}$ and sin$^2\theta_{23}$ ending up approximately $3\sigma$ larger than the observed values; sin$^2 \theta_{13}$ and $\delta_{CP}$ deviate less than $1\sigma$ from the observed values. Fig. \ref{fig:dmix} shows mixing angle distributions calculated at uniform sampling within a limited box of the four-dimensional parameter space. Points within $1\sigma$ from the experimental sin$^2 \theta_{13}$ and $\delta_{CP}$ are red colored. The best fit is obtained for the red points closest to the intersection of the shaded areas in the left-most subplot. Higher-order groups, such as $C_i\times S_4$ and $C_4\times A_4$ give similar distributions.  

\section{Conclusions}
We have presented a method to determine discrete symmetries for the Yukawa sector in the 3HDM model, which gives the relationship between the masses of leptons and the elements of the PMNS mixing matrix.
The implied relations among the lepton masses and mixing can be compared to the experimental data. Using the method
we have searched for discrete flavour symmetry groups with order less than 1035. With the applied representation selection criteria, none of the studied groups provides a symmetry which predicts lepton masses and mixing simultaneously compatible with the experimental data. Partial and approximate agreement with the data has been explored and appeared to exist for a small number of groups.

The method can accommodate different selection criteria by altering the presented algorithm. For example the restriction to $U(3)$ subgroups can be relaxed, and unfaithful, reducible and/or lower-dimensional representations can be admitted. Also, the requirement that the tensor product decomposition is unique could be omitted, allowing multi-dimensional solutions. Presumably it will thus be feasible for a given group, to either detect a symmetry compatible with the experimental data or to rule out any such symmetry even for the most relaxed hypothesis.

There are several possible improvements and extensions to the described method.
The present results are based on numerical sampling in four-dimensional
VEV-space, where the choice of search interval and the sampling density is
limited for practical reasons. The analysis would be highly enhanced
when analytic expressions for the eigenvalues of mass matrices are used
to find bounds on solutions of physical quantities.

\funding{This work has been supported in part by the Polish National Science Center (NCN) under grant 2020/37/B/ST2/02371 and the Research Excellence Initiative of the University of Silesia in Katowice.}

\dataavailability{The data presented in this study are available upon reasonable request
from the corresponding authors.} 

\conflictsofinterest{The authors declare no conflict of interest.} 
\authorcontributions{All authors contributed equally to the reported research. All authors have read and agreed
to the published version of the manuscript.}
\acknowledgments{
We thank Jacek Holeczek for his help. We are very grateful to the GAP Support Group for their advising.}
\bibliography{Main}
\end{document}